\documentclass[a4paper,twocolumn, altadd, nofootinbib,superscriptaddress,aps,prb,citeautoscript, 14pt,
eqsecnum,notitlepage,showkeys]{revtex4-1}
\usepackage{multirow}
\usepackage{graphicx}
\usepackage{bm}
\usepackage{dcolumn}
\usepackage[hidelinks]{hyperref}
\usepackage{gensymb}
\usepackage{amsmath}
\usepackage{dcolumn}    
\usepackage{ifpdf}
\usepackage{amssymb,lineno,amsfonts}
\usepackage{graphicx}   
\usepackage{bm}         
\usepackage{bbm}
\usepackage{mathrsfs}
\usepackage{upgreek}
\usepackage{mathtools}
\usepackage{epstopdf}
\usepackage{setspace}
\usepackage{hyperref}
\usepackage{natbib}
\usepackage{esvect}
\usepackage[usenames,dvipsnames]{xcolor}
\definecolor{med-blue}{RGB}{25,25,112}
\hypersetup{colorlinks, linkcolor={blue},citecolor={blue}, urlcolor={blue}}
\usepackage{times	}
\usepackage [english]{babel}
\usepackage [autostyle, english = american]{csquotes}
\MakeOuterQuote{"}

\begin{document}
\title{A reentrant superspin glass state and magnetization steps in the oxyborate Co${_2}$AlBO$_{5}$}
\author{Jitender Kumar}
\affiliation{Department of Physics, Indian Institute of Science Education and Research, Dr. Homi Bhabha Road, Pune 411008, India}
\author{Soumendra Nath Panja}
\affiliation{Department of Physics, Indian Institute of Science Education and Research, Dr. Homi Bhabha Road, Pune 411008, India}
\author{Deepak John}
\affiliation{Department of Physics, Indian Institute of Science Education and Research, Dr. Homi Bhabha Road, Pune 411008, India}
\author{Arpan Bhattacharyya}
\affiliation{Saha Institute of Nuclear Physics, 1/AF Bidhannagar, Kolkata, India }
\author{A. K. Nigam}
\affiliation{Department of Condensed Matter Physics and Material Science, Tata Institute of Fundamental Research, Dr. Homi Bhabha Road, Mumbai 400 005, India}
\author{Sunil Nair}
\affiliation{Department of Physics, Indian Institute of Science Education and Research, Dr. Homi Bhabha Road, Pune 411008, India}
\affiliation{Centre for Energy Science, Indian Institute of Science Education and Research, Dr. Homi Bhabha Road, Pune 411008, India}
\date{\today}
\begin{abstract} 
An oxyborate Co${_2}$AlBO{$_{5}$} belonging to the ludwigite family is investigated using structural, thermodynamic, dielectric and magnetic measurements. Magnetic measurements indicate that this system is seen to exhibit long range magnetic ordering at $T{_N}$ = 42 K, signatures of which are also seen in the specific heat, dielectric susceptibility, and the lattice parameters. The absence of a structural phase transition down to the lowest measured temperatures, distinguishes it from the more extensively investigated Fe-based ludwigites. At low temperatures, the system is seen to stabilize in a reentrant superspin glass phase at $T{_G} =$ 10.6 K from within the magnetically ordered state. This ground state is also characterized by magnetic field induced metamagnetic transitions, which at the lowest measured temperatures exhibit a number of sharp magnetization steps, reminiscent of that observed in the mixed valent manganites. 
\end{abstract}
\pacs{Pacs}
\keywords{Keywords}
\maketitle
\section{Introduction}

The anisotropy inherent to different structural motifs play a critical role in determining the electronic and magnetic ground states of a number of strongly correlated electron systems. The complex interplay between the spin, charge and lattice degrees of freedom in such restricted geometries are known to give rise to phenomena ranging from unconventional superconductivity, to novel spin and charge density waves.  Oxyborates of the form $M{_2}M'BO{_5}$ crystallizing in the \emph{Ludwigite} structure constitutes one such family of strongly correlated oxides, which have attracted interest in the recent years \cite{Fe1, Fe2, Fe3, Co1}. Here, $M$ and $M'$ are divalent and trivalent metals respectively, which drive the electronic and magnetic properties of these systems. Systems in which $M$ and $M'$ are made up of the same element are called Homometallic Ludwigites, with Co${_3}$BO${_5}$ and Fe${_3}$BO${_5}$ being the most well-investigated examples. Though both of them crystallize in the same orthorhombic symmetry, the magnetic and structural properties vary considerably \cite{Co1, Co2, Co3}. For instance, on cooling from room temperature, Fe${_3}$BO${_5}$ first exhibits a structural transition (at $\approx$ 280 K), followed by a couple of low temperature magnetic transitions (at 112 K and 70 K) associated with the progressive ordering of Fe on different crystallographic sites. On the other hand, Co${_3}$BO${_5}$ is reported to exhibit a single magnetic transition at $\approx$ 45 K. Though the interactions are predominantly antiferromagnetic, static and dynamic susceptibility measurements on both these systems indicate that a weak ferromagnetic component exists along with long range antiferromagnetic order. The Hetrometallic Ludwigites, where $M$ and $M'$ are made up of different elements, can be broadly divided into two types. In some systems, both $M$ and $M'$ are magnetic (for instance, $M=$ Co, Ni, Mn, or Cu, and $M'=$ Fe, Ni or Cr) and both these species appear to participate in the observed magnetic order \cite{mag1, mag2, mag3}. The other type comprises of systems in which $M'$ is non-magnetic (for instance, Ga, Al, Mg, or Ti), and the resultant dilution of the magnetic lattice typically manifests itself in the form of a reduced transition temperature \cite{dil1, dil2, dil3, dil4}. Since the $M$ ions has four in-equivalent crystallographic sites within the Ludwigite structure, the eventual properties are likely to be crucially influenced by the occupancies of the different $M$ and $M'$ species within these sites. In some Ludwigites, this is also reported to give rise to a low temperature spin glass like state without long range order \cite{dil2, dil4}. 

\begin{figure}
   \vspace*{0.6cm}
	\hspace*{-0.1cm}
	\includegraphics[scale=0.35]{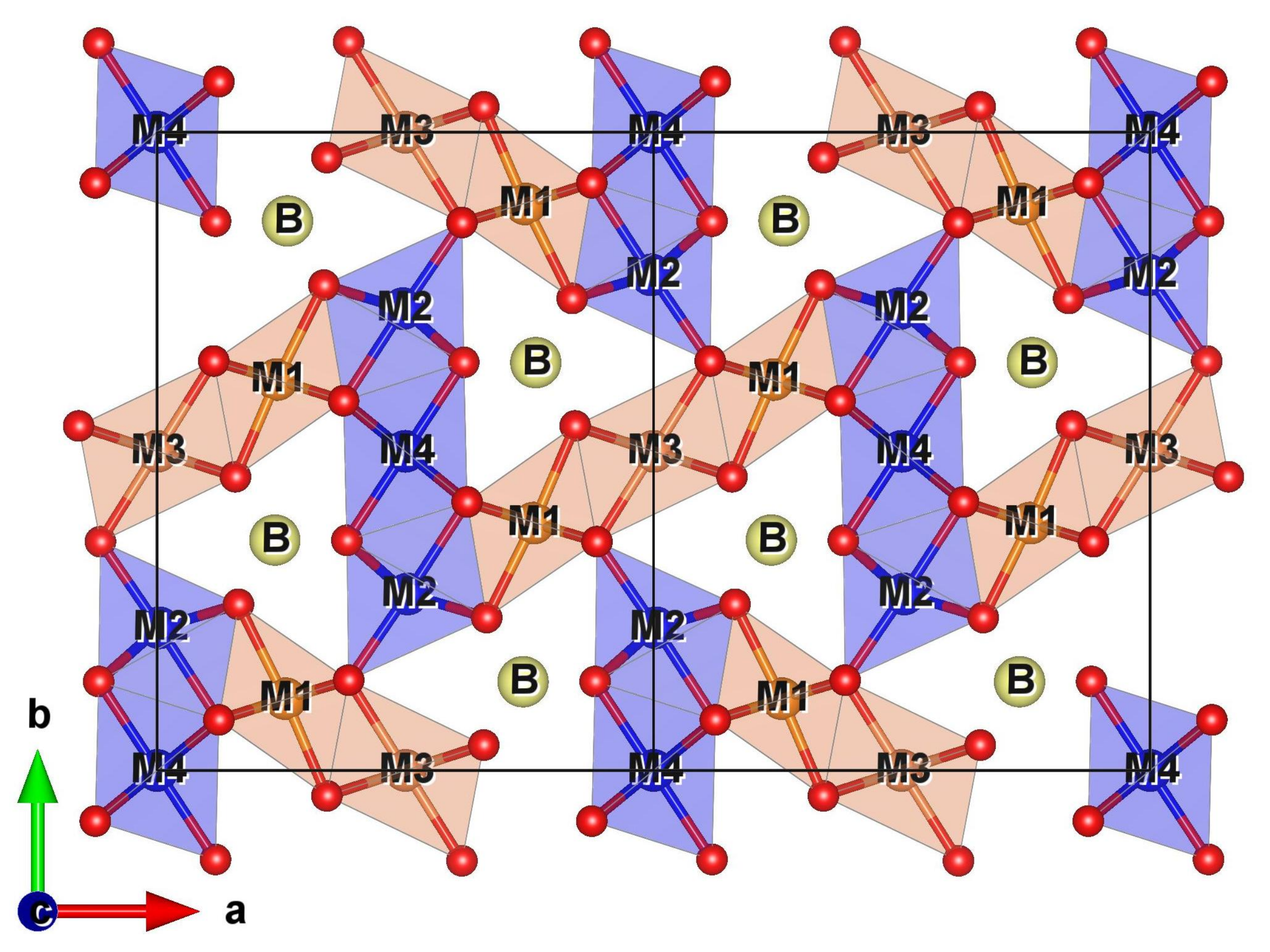}
	\caption{The crystal structure of the homometallic ludwigite M${_3}$BO${_5}$ as viewed along the crystallographic $c$ axis. $M1$ to $M4$ depict the four distinct crystallographic positions which the transition metal can occupy within this structure. }
	\label{figure1}
\end{figure}

A defining feature of the Ludwigite structure is the presence of a zig-zag wall made up of edge sharing octahedra which propagate along the crystallographic $c$ axis (Figure \ref{figure1}). 
The four possible crystallographic positions of the $M$ ion are thus distinguished by their positions within these zig-zag structures. In the Fe${_3}$BO${_5}$ system, the ions within the crystallographic site 2 and 4 are involved in a structural transition at $\approx$ 280 K. Driven by a displacement of the central Fe${^2}$ ion in alternate directions along the Fe${^4}$-Fe${^2}$-Fe${^4}$ triad, this effectively doubles the lattice period along the crystallographic $c$ axis \cite{Fe3}. Interestingly, no such structural transition is reported either in the closely related Co${_3}$BO${_5}$ system, or in any other heterometallic ludwigite.  

Here, we report on the hitherto unexplored Co${_2}$AlBO$_{5}$ system, and investigate it using temperature dependent x-ray diffraction, specific heat, magnetization and dielectric measurements. We observe the presence of a low temperature re-entrant $superspin$ glass like state from within an (antiferro)magnetically ordered phase. Within this phase, a number of field induced transitions are observed, which at the lowest measured temperatures manifests itself in the form of sharp magnetization  steps.  

\section{Experimental Techniques}

Small needle like crystallites of Co${_2}$AlBO$_{5}$ were synthesized using a reactive flux technique. Stoichiometric amounts of Co${_3}$O${_4}$ and Al${_2}$O${_3}$ were mixed with excess borax (Na${_2}$B${_4}$O${_7}$. 10H${_2}$O) in the ratio of 1:5, and ground well using ball mill at 120 rpm for 12 hours to make a fine homogeneous mixture. This mixture was treated at 1000$\degree$C for 90 hours in an alumina crucible followed by slow cooling to 740$\degree$C (at the rate of 5$\degree$C/hour), after which the furnace was turned off. Fine needle-like crystallites of the target material was extracted from the crucible, and was washed using warm dilute HCl and distilled water to remove the excess borax. Since these crystallites were too small for routine magnetic and thermodynamic measurements, they were crushed and treated as polycrystalline powders. Phase purity was confirmed using a Bruker D8 Advance diffractometer with Cu K$_{\alpha}$ radiation. Low temperature X-Ray diffraction was obtained using the  powder diffractometer at beamline BL-18B, Photon Factory, KEK, Japan using a x-ray wavelength of  0.8019 ${\AA}$. Room temperature structural details were rigorously analyzed by the Reitveld method using the Fullproof  refinement program \cite{Fullprof}, and the variation in the lattice parameters of the low temperature scans were determined using a Le Bail fit. Elemental compositions and their homogeneity were reconfirmed by using an energy dispersive X-Ray spectrometer (Ziess Ultra Plus).  Specific heat and magnetization measurements were performed using a Quantum Design PPMS and a MPMS-XL SQUID magnetometer respectively. Temperature dependent dielectric measurements were performed in the standard parallel plate geometry, using a HP 4294A Impedance Analyzer. Measurements were done using an excitation ac signal of 1V. 

\section{Results and Discussions}
\begin{figure}
	\vspace*{-0.2cm}
	\hspace*{-0.1cm}
\includegraphics{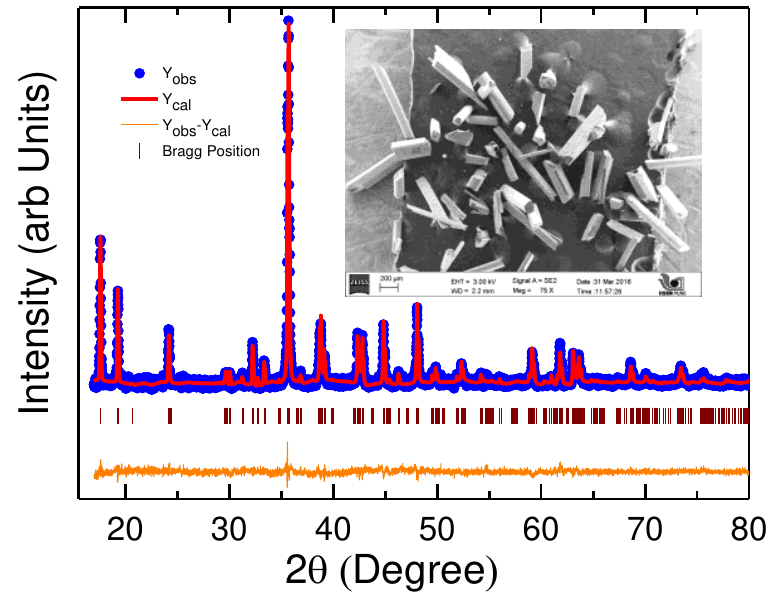}
\caption{A Rietveld fit to the room temperature x-ray diffraction data of Co${_2}$AlBO$_{5}$. This corresponds to a fit with $R$ parameters of $R{_{wp}}$ = 15.0, $R{_e}$ = 12.0, and $\chi{^2}$ = 1.56. The inset shows a scanning electron micrograph where long needle like crystallites are seen.}
\label{figure2}
\end{figure}

Room temperature X-ray diffraction of Co${_2}$AlBO$_{5}$ revealed a single phase specimen, with no trace of any of the starting materials or any other impurity phases. A Scanning Electron micrograph reveals long needle-like crystallites as is shown in the inset of Figure \ref{figure2}. A Rietveld refinement of the room temperature diffraction data is shown in the main panel of Figure \ref{figure2}. 
\begin{table}
	
	\begin{tabular}{|c|c|c|c|c|c|}
		\hline
		\multicolumn{6}{|c|} {Co${_2}$AlBO$_{5}$ } \\
		\multicolumn{6}{|c|}{Temperature = 296 K} \\
		\multicolumn{6}{|c|}{Space Group : $Pbam$} \\
		\multicolumn{6}{|c|}{Crystal system: Orthorhombic }\\
		\multicolumn{6}{|c|}{ a= 12.034(9) ${\AA}$}\\
		\multicolumn{6}{|c|}{ b=9.200(9) ${\AA}$ }\\
		\multicolumn{6}{|c|}{c= 2.997(8) ${\AA}$ }\\
		\multicolumn{6}{|c|}{${\alpha}$ = ${\beta}$ = ${\gamma}$ = 90 \degree} \\
		\hline
		\hline
		Atom & Wyckoff & x/a &y/b & z/c & Occupancy\\
		\hline
		Co$^1$  & 4h & 0.2752(2) & 0.9985(7)  & 0.5 & 0.6916\\
		Al$^1$  & 4h & 0.2752(2) & 0.9985(7)  & 0.5 & 0.3084\\
		Co$^2$  & 2b & 0 & 0 & 0.5 & 0.865\\
		Al$^2$  & 2b & 0 & 0 & 0.5 & 0.135\\
		Co$^3$  & 2c & 0 & 0.5 & 0 & 0.549\\
		Al$^3$  & 2c & 0 & 0.5 & 0 & 0.451\\
		Co$^4$  & 4g & 0.1144(5) & 0.2357(5)  & 0 & 0.458\\
		Al$^4$  & 4g & 0.1144(5) & 0.2357(5)  & 0 & 0.542\\
		B$^1$  & 4g & 0.373(4)  & 0.256(3)  & 0 & 1 \\
		O$^1$  & 4g & 0.4596(10) & 0.3610(12)&  0 & 1 \\
		O$^2$  & 4h & 0.1354(14) & 0.1204(10) & 0.5 & 1\\
		O$^3$  & 4g & 0.3572(14) & 0.1284(10) & 0 & 1 \\
		O$^4$  & 4h & 0.0735(11) & 0.3972(9) & 0.5 & 1\\
		O$^5$  & 4g & 0.2576(11) & 0.3328(13) & 0 & 1 \\
		\hline
	\end{tabular}
	\caption{Structural Parameters of Co${_2}$AlO${_2}$BO$_{5}$ as determined from the Rietveld analysis of room temperature X-ray diffraction data. }
	\label{Table1}
\end{table}
This system was seen to crystallize in an orthorhombic $Pbam$ symmetry, and the structural details of Co${_2}$AlBO$_{5}$ as determined from the Rietveld refinement of room temperature X-ray diffraction data is summarized in Table \ref{Table1} . All the 4 crystallographic sites available for Co are seen to be diluted with the non-magnetic Al, which is in contrast to that reported in the related Co${_{2.4}}$Ga${_{0.6}}$BO${_5}$ system, where it was reported that Ga does not occupy the Co$^1$ and Co$^3$ sites \cite{dil1}. However, in our case as well, the Co$^4$ and Co$^2$ sites appear to be the most preferred and least preferred sites respectively. It is to be noted that in the Co${_2}$AlBO$_{5}$ system, $a > b$, whereas this is reversed ( ie. $a < b$) in the parent Co${_3}$BO$_{5}$, the closely related Fe${_3}$BO$_{5}$, as well as all of their doped variants reported till date. Interestingly, this appears to be a feature unique to the Co and Ni aluminoborates \cite{hriljac}. 

\begin{figure}
	\vspace*{-0.9cm}
	\hspace*{-1.0cm}
	\includegraphics[scale=0.36]{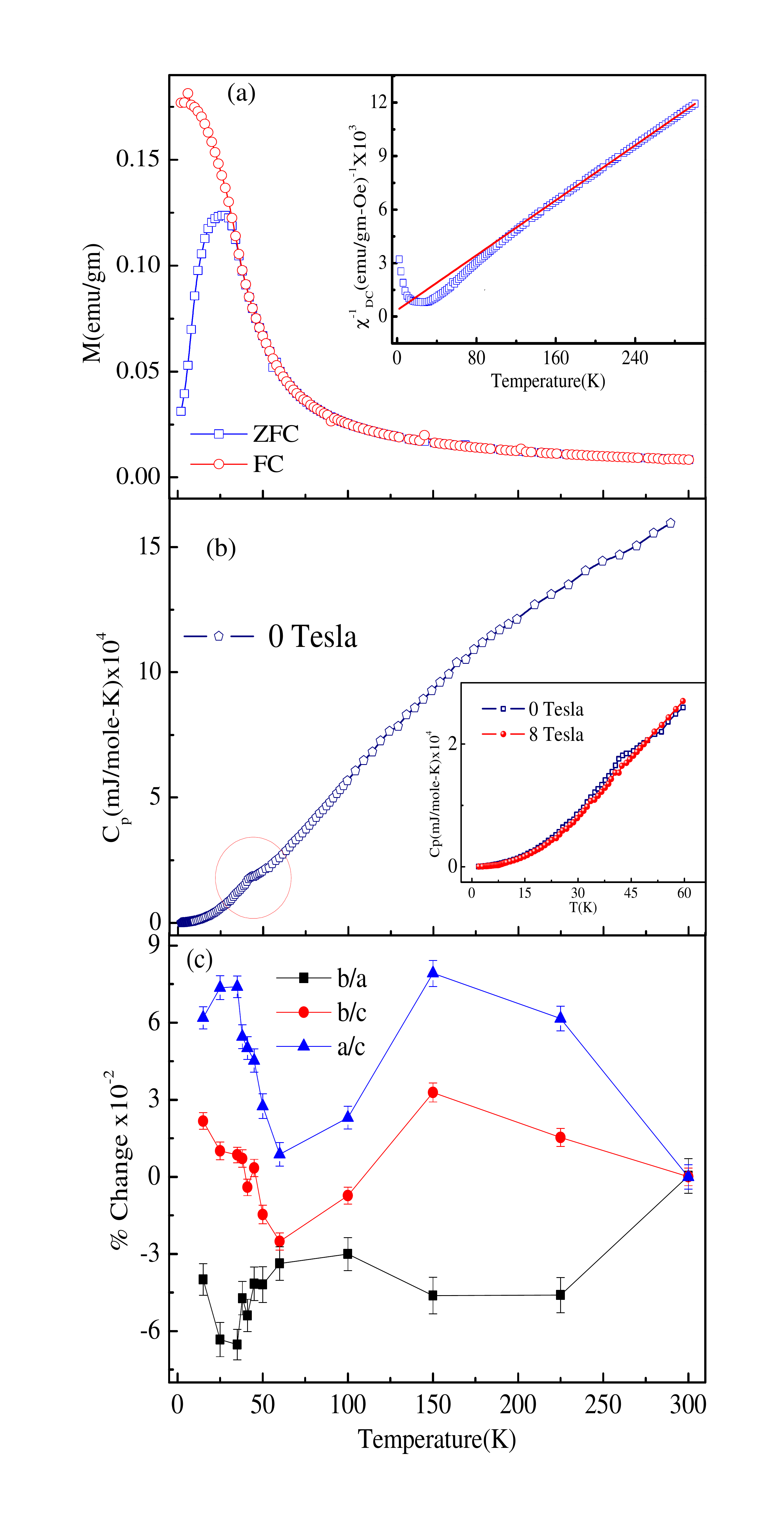}
	\caption{(a) depicts the temperature dependence of dc magnetization as measured in Co${_2}$AlBO$_{5}$ in the Zero Field Cooled (ZFC) and Field Cooled (FC) protocols. The inset shows a Curie Weiss fit in the high temperature paramagnetic region. (b) depicts the heat capacity as measured in the same sample in zero applied field, with a peak at $\approx$ 42 K indicating the onset of magnetic order. The inset shows the suppression of this feature with an applied magnetic field of 8 Tesla. (c) depicts the $\%$ change in the lattice parameter ratios ($b/a$, $b/c$ and $a/c$ normalized with respect to their room temperature values), with a clear change being observed in the phase transition region.}
	\label{figure3}
\end{figure}

The parent Co${_3}$BO${_5}$ system is reported to exhibit long range antiferromagnetic order  and diluting the magnetic lattice using nonmagnetic Al would be expected to reduce temperature where long range order sets in. Figure  \ref{figure3}(a) shows the temperature dependence of the dc magnetization as measured in the Co${_2}$AlBO$_{5}$ specimen. A bell shaped feature, with a pronounced splitting between the Zero Field Cooled (ZFC) and Field Cooled (FC) measuring protocols is observed.The temperature dependence of specific heat exhibits a small feature at around 42 K, which possibly corresponds to the onset of long range magnetic ordering (Figure \ref{figure3} (b) ). This also broadly coincides with the temperature at which the ZFC and FC curves bifurcate. Application of magnetic fields of the order of 8 Tesla is seen to smear off this feature in the specific heat, as is seen in the inset. A Curie Weiss fit to the inverse dc magnetic susceptibility in the high temperature paramagnetic region as shown in the inset of Figure  \ref{figure3}(a) gives a Curie-Weiss temperature $\theta{_{CW}}= -9.19K$. In comparison, the $\theta{_{CW}}$ values for Fe${_3}$BO${_5}$ and Co${_3}$BO${_5}$ are reported to be -485K and -25K respectively \cite{Fe3,Co1}. It is evident that the variation of $\theta{_{CW}}$ in these ludwigites are not consistent with the temperatures at which long range magnetic order is observed, since the ratio $\left|\theta{_{CW}}/T{_N}\right|$ is 6.9 and 0.6 for Fe${_3}$BO${_5}$ and Co${_3}$BO${_5}$ respectively. This anomalously small $\left|\theta{_{CW}}/T{_N}\right|$ value in the Co- system was attributed to the predominance of ferromagnetic interactions, since ferromagnetically ordered rungs were thought to align anti-parallely at the global $T{_N}$ \cite{Co1}. This was also corroborated by larger values of the spontaneous magnetization determined from MH isotherms within the magnetically ordered state.  The  Co${_2}$AlBO$_{5}$ specimen investigated here exhibits a $\left|\theta{_{CW}}/T{_N}\right|$ value of 0.21, which is even smaller than that observed for the Co${_3}$BO${_5}$ system. This could be a consequence of the fact that nonmagnetic Al dilutes the magnetic sublattice.  This could also indicate that Al substitution appears to favour the stabilization of ferromagnetic interactions at the cost of the antiferromagnetic ones. The effective magnetic moment per Co${^{2+}}$ ion calculated from the Curie-Weiss fit is 3.5$\mu{_{\beta}}$, which is close to the high spin value of Co${^{2+}}$. 
Crystallizing in similar orthorhombic ($Pbam$) structures at room temperatures, a striking difference between the Fe and the Co based ludwigites is the observation of a temperature driven structural phase transition in the former, where a $Pbam$$\rightarrow$$Pbnm$ symmetry change occurs at $\approx$ 280K \cite{Fe3}. On the contrary, the Co based system has shown no evidence of such a transition at least down to $\approx$ 100K \cite{Co1}, though it remains to be investigated whether such a transition exists at lower temperatures. Our temperature dependent x-ray diffraction measurements on the Co${_2}$AlBO$_{5}$ system down to 15K clearly rule out the presence of a structural transition in this system, indicating that the absence of a structural transition appears to be a defining feature of all the Co based ludwigites. However our measurements also indicate subtle changes in the lattice parameters as a function of temperature and the effective change in the ratios $b/c$, $a/c$ and $b/a$ is shown in Figure 3c. Two distinct inflection points are seen - one at $\approx$ 150 K, which is not discernible in any other measurement, and the other at $\approx$ 50K, which appears to coincide with the onset of long range magnetic order. The latter indicates that the lattice and spin degrees of freedom could be strongly coupled in this system.   

Measurements of the dielectric constant in magnetic materials offer an interesting avenue of evaluating the coupling between the charge and magnetic degrees of freedom. To the best of our knowledge, there have been no reports of dielectric measurements of any member of the Ludwigite family. Figure 4a depicts the real part of the dielectric constant as a function of temperature, as measured at 685 kHz. 
\begin{figure}
		\vspace*{-1.08cm}
		\hspace*{-1.5cm}
	\includegraphics[scale=0.35]{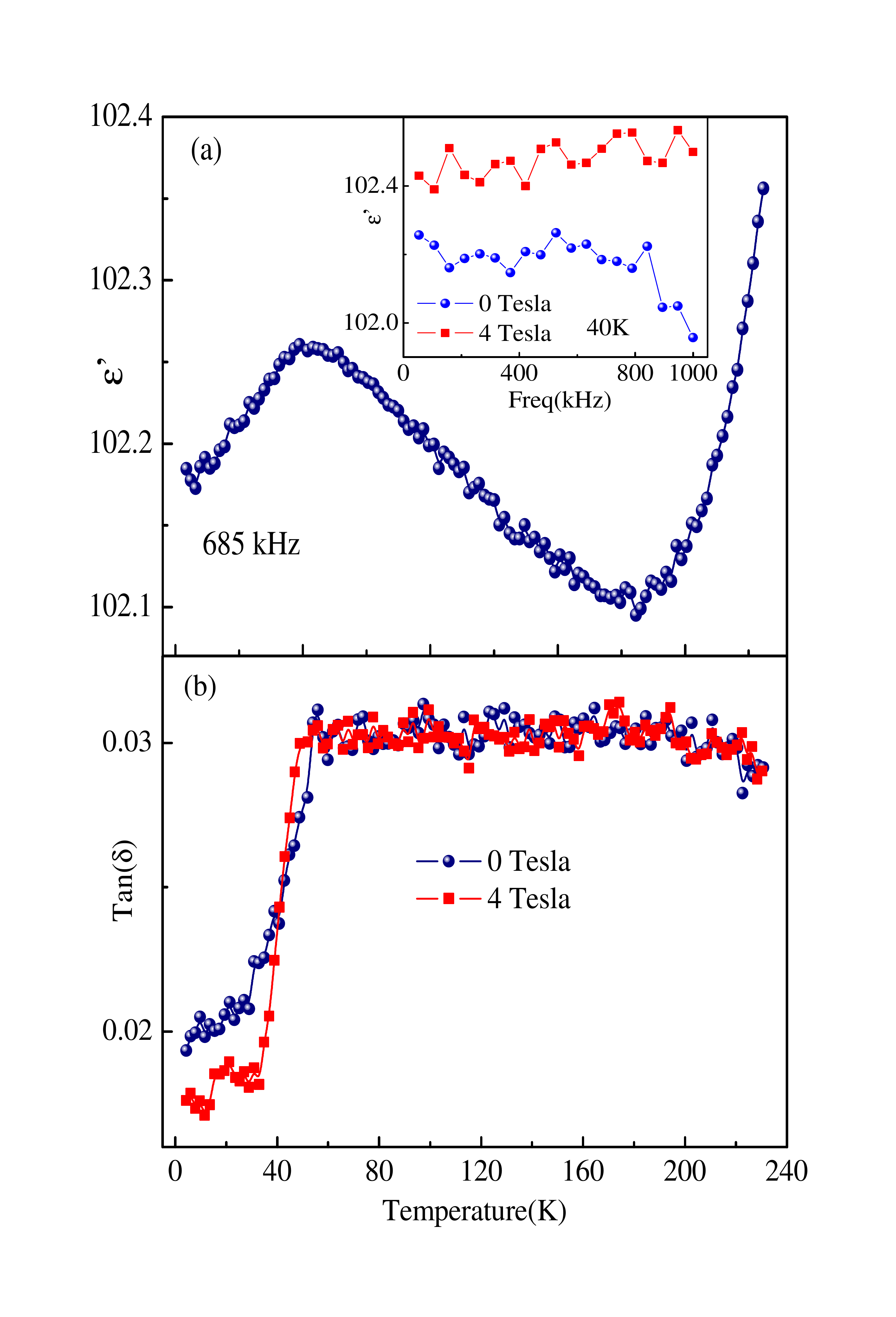}
	\caption{(a) depicts the temperature dependence of the real part of the dielectric constant $\epsilon' (T)$ (measured at a frequency of 685 kHz and an excitation voltage of 1 V${_{ac}}$) which peaks at the magnetic ordering temperature. The inset shows a finite magnetoelectric effect as measured in isotherms of the dielectric constant $\epsilon' (H)$ at $T =$ 40 K. (b) shows the dielectric loss as measured at 0 and 4 Tesla, with both of them exhibiting a sharp feature across the magnetic transition.}
	\label{figure4}
\end{figure}
\begin{figure}
	\vspace*{-0.9cm}
	\hspace*{-1.1cm}
	\includegraphics[scale=0.37]{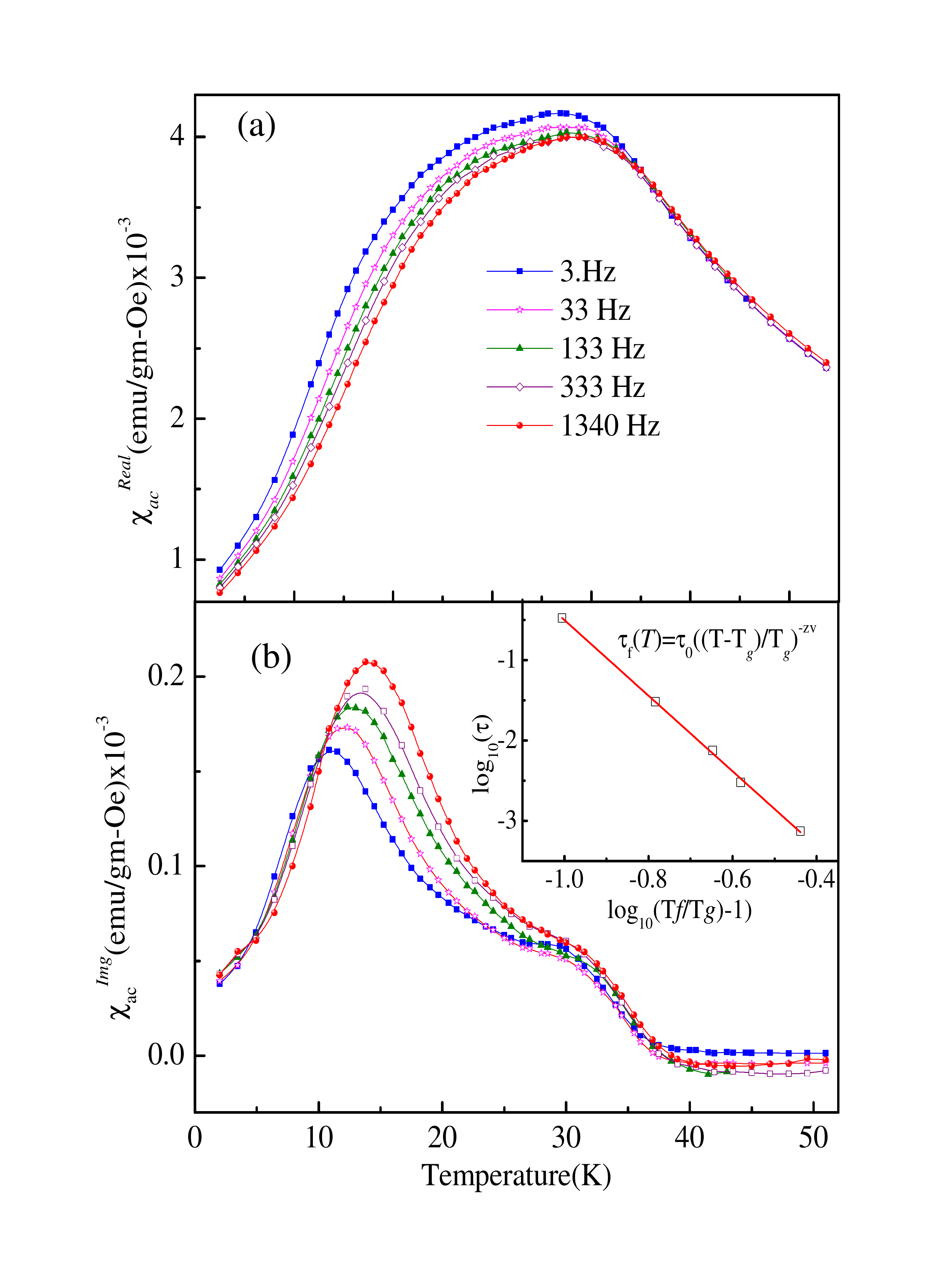}
	\caption{The real (a) and the imaginary (b) parts of the ac magnetic susceptibility as measured in Co${_2}$AlBO$_{5}$. The presence of a frequency dependent low temperature transition and a frequency independent high temperature one is clearly seen. The inset of (b) depicts dynamical scaling using $\tau/\tau{_0} = ((T{_f}-T{_G})/T{_G})){^{z\nu}}$, with the best fit giving  $T{_G}$ = 10.6 $\pm0.2$ K and $z\nu$ = 4.7 $\pm0.5$.}
	\label{figure5}
\end{figure}
A maximum in $\epsilon'(T)$ is observed in the vicinity of the magnetic transition, indicating a coupling of the electric and magnetic order parameters in this system. To further investigate this phenomenon, we have measured isotherms of the dielectric constant under zero and 4 Tesla magnetic field within the magnetically ordered state, as is shown in the inset. A finite difference ($\approx $0.6 $\%$) is observed, with $\epsilon' (H)$ being larger in the presence of an applied field. The magneto-dielectric nature of Co${_2}$AlBO$_{5}$ is also evident from the dielectric loss data (Figure Figure  \ref{figure4}(b)), where a clear drop is observed in both the zero field and in-field data near the magnetic ordering temperature.

It is to be noted that there have been conflicting reports on the nature of the low temperature magnetic ground state in the parent Co${_3}$BO${_5}$ system. Ivonava and co-workers \cite{Co2} have reported a long range magnetic transition with a $T{_N}$$\approx$ 45K, and a second magnetic transition at $\approx$17K, as evidenced from a drop in the dc magnetization, coupled to a bifurcation of the ZFC and FC measurements. A subsequent publication \cite{Co1} however reported only a solitary high temperature transition at $T{_N} \approx$ 42 K. Though ac susceptibility measurements within the magnetically ordered phase was highly frequency dependent, no clear signature of an additional low temperature transition was observed in either the magnetic or thermodynamic measurements. It has also been suggested that the transition at $\approx$ 42K is predominantly ferromagnetic, with an overriding weak antiferromagnetic component. In the absence of conclusive neutron diffraction data, the true magnetic structure of the Co${_3}$BO${_5}$ system remains undetermined. To investigate the possibility of an additional low temperature transition in Co${_2}$AlBO$_{5}$, we have performed frequency dependent ac susceptibility measurements, as is shown in Figure. \ref{figure5} .

 As seen in the upper panel, though a finite dispersion is observed at low temperatures in the real part of the ac magnetic susceptibility ($\chi{{^R}_{ac}}$), there is no clear evidence of an additional low temperature transition. However, in the imaginary part of the susceptibility ($\chi{{^I}_{ac}}$), two clear transitions are observed - a frequency independent high temperature transition, and a low temperature frequency dependent one. The former is clearly associated with the onset of long range (antiferro / ferri)-magnetic order, with the latter characterizing the presence of a reentrant glassy phase. 

The stabilization of a reentrant glassy phase from within a state with magnetic order is well known, and is known to exist in a number of different material classes including intermetallics \cite{inter}, oxides \cite{anil} and interacting nanoparticle \cite{nano} systems. Unlike in a prototypical paramagnetic - spin glass transition, where disorder or mixed exchange interactions gives rise to an \emph{atomistic} glassy phase with frozen spins, these reentrant glasses are characterized by the freezing of \emph{superspins} which could have ferromagnetic, antiferromagnetic or ferrimagnetic order within. It has been unambiguously demonstrated that such systems exhibit a dynamic behavior analogous to that of prototypical spin glasses \cite{mat}. The inset of Figure 5b shows a fit to a conventional critical slowing of the relaxation times given by $\tau/\tau{_0} = ((T{_f}-T{_G})/T{_G})){^{z\nu}}$. Here, $T{_f}(\omega)$ and $T{_G}$ refers to the freezing temperature characterized by the maximum in the $\chi{{^I}_{ac}}$, and the true reentrant glass transition temperature respectively.  The best fit of the data, spanning from 3 Hz to 1.3 kHz yields a glass transition temperature $T{_G}$ = 10.6 $\pm0.2$ K, a critical exponent $z\nu$ = 4.7 $\pm0.5$, and an effective spin flip time $\tau{_0}$ of 10${^{{-6}}}$ seconds. We note that the value of $\tau{_0}$ in most atomistic spin glasses are of the order of $10{^{-12}} - 10{^{-13}}$ seconds \cite{atom1,atom2,atom3}, whereas in the case of interacting superparamagnets, values of the order of $10{^{-6}} - 10{^{-9}}$ seconds have been reported \cite{super1,super2,super3}. The fact that $\tau{_0} \approx 10{^{{-6}}}$ seconds in our case is clearly a consequence of the fact that the magnetic entity under consideration here is made up of a large number of spins. We also note that the critical exponent $z\nu$ in our case appears to be smaller than that observed in most spin glass transitions. 

\begin{figure}
		\vspace*{-0.9cm}
		\hspace*{-1.0cm}
	\includegraphics[scale=0.36]{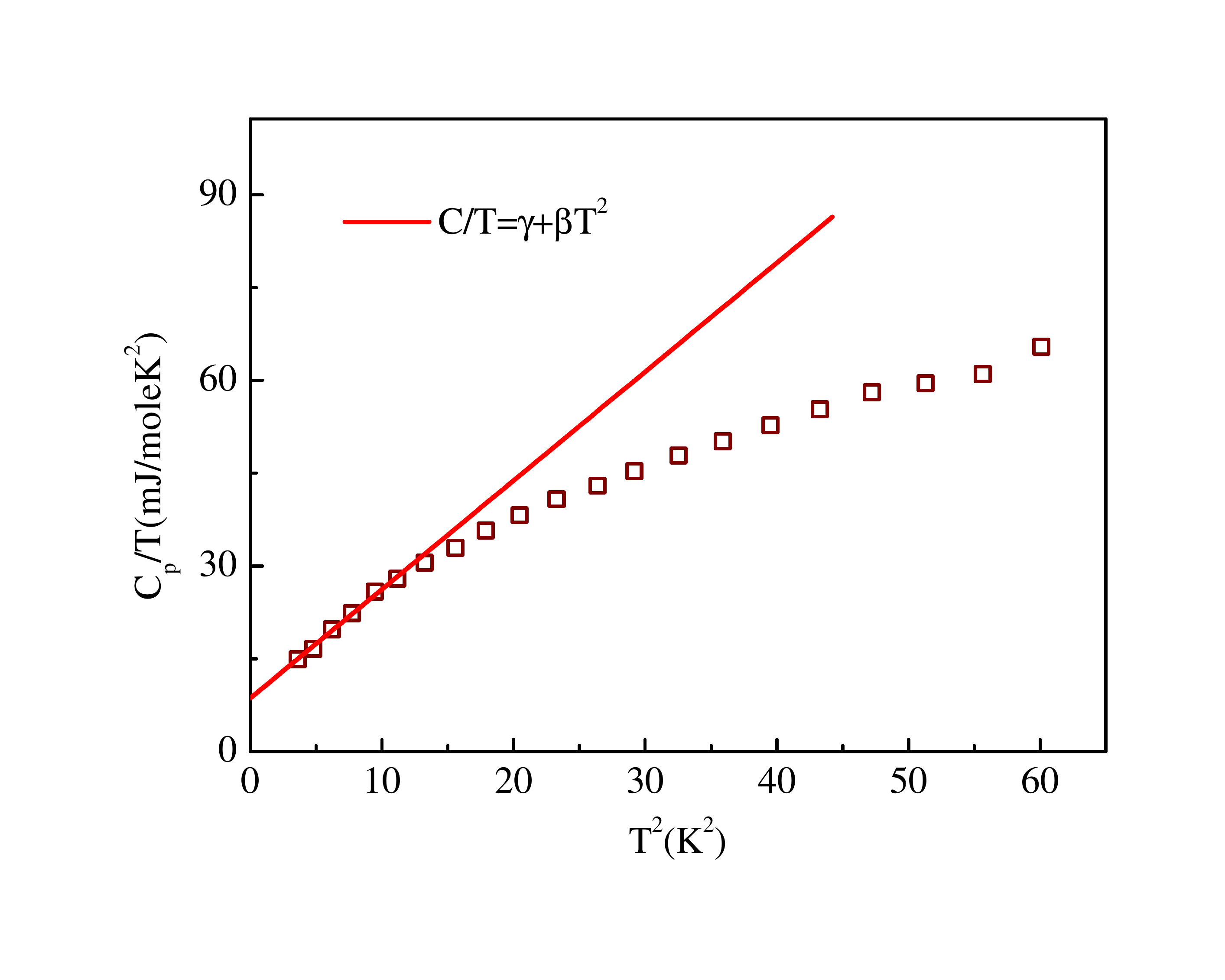}
	\caption{The specific heat of Co${_2}$AlBO$_{5}$ plotted as $C/T$ versus $T{^2}$, with the linear fit for T $<$ 4 K indicating that the data can be described by the equation $C/T$ = $\gamma$+$\beta T{^2}$. The values of $\gamma$ and $\beta$ obtained from this fit are 8.6 $\pm0.6$ mJ/mol K${^2}$ and 1.75 $\pm0.07$ mJ/mol K${^2}$ respectively. }
	\label{figure6}
\end{figure}

The glassy nature of the low temperature magnetic ground state is also reflected in the analysis of the specific heat measurements. Figure \ref{figure6} shows the linear fit to the low temperature specific heat using the equation $C/T = \gamma + \beta T{^2}$, giving values of 8.6 $\pm0.6$ mJ/mol K${^2}$ and 1.75 $\pm0.07$ mJ/mol K${^2}$  for $\gamma$ and $\beta$ respectively. The large value of $\gamma$ in the ludwigites has earlier been attributed to magnetic frustration in these materials \cite{sp_ht}. It is to be noted that the value of $\gamma$ in the case of Co${_2}$AlBO$_{5}$ is more than twice as large as that reported for the undiluted homo-metallic Co based ludwigite . 
The only other ludwigite with a larger $\gamma$ is the system Co${_5}$Ti(O${_2}$BO${_3}$)${_2}$ which was reported to stabilize in a low temperature spin glass phase, exhibiting a conventional paramagnetic - spin glass transition at $T{_G}$ $\approx$ 19 K \cite{SG}. The value of the Debye temperature ($\theta{_D}$) using the relation $\theta{_D}{^3}$ = 243$R/\beta$ (with $R$ being the universal gas constant) give a value of 104.7 K, which is smaller than that determined earlier for the undiluted Co${_3}$BO${_5}$ system.  

\begin{figure}
		\vspace*{-0.9cm}
		\hspace*{-1.0cm}
	\includegraphics[scale=0.36]{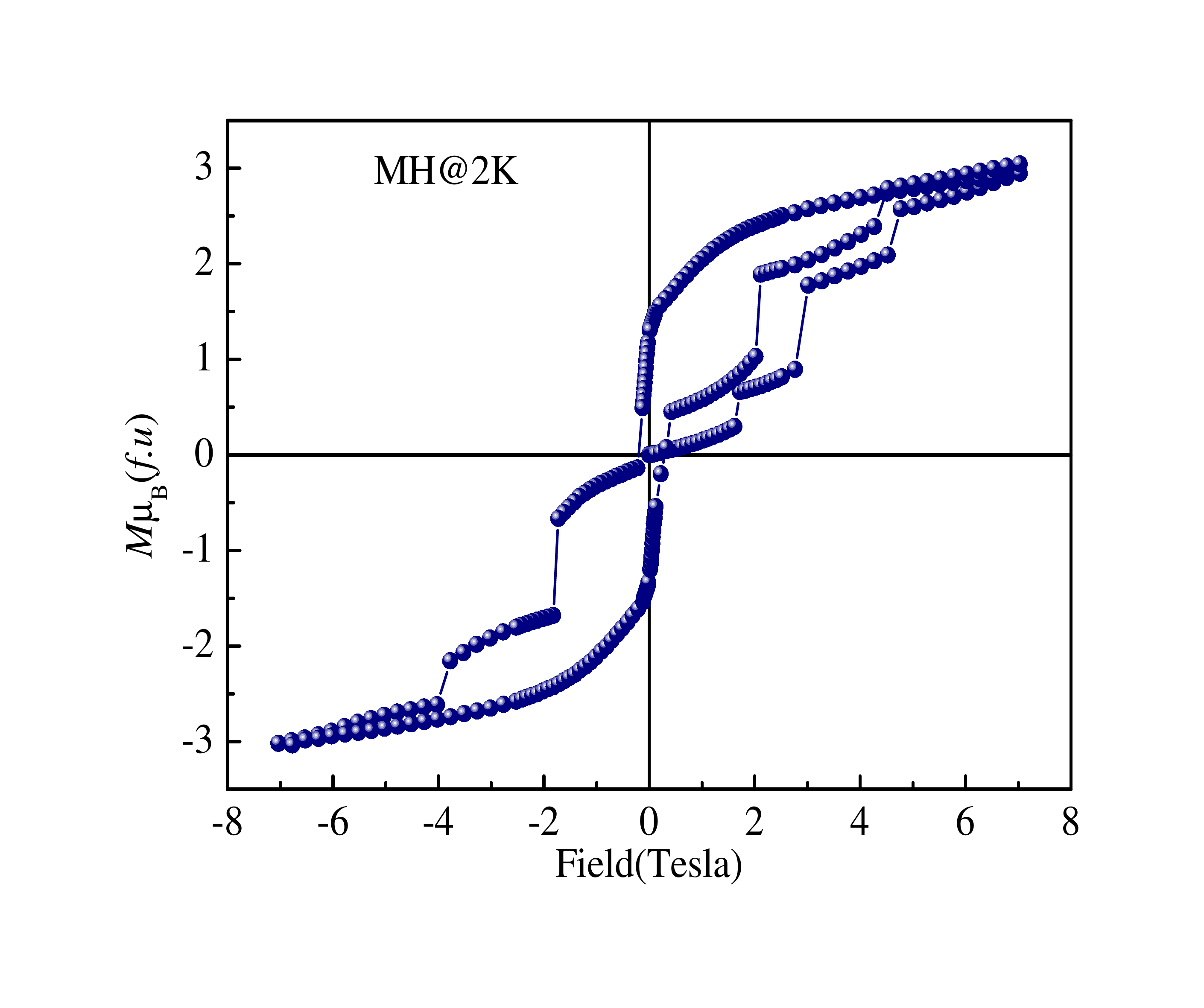}
	\caption{The magnetization isotherm of Co${_2}$AlBO$_{5}$ as measured at 2 K, exhibiting a number of sharp magnetization steps.}
	\label{figure7}
\end{figure}

The Co${_3}$BO${_5}$ system is reported to exhibit an uniaxial anisotropy, with the easy direction of magnetization lying along the crystallographic $b$ axis \cite{mag2}. Below $T \leq 20K$, a stiffening of the hardness has also been suggested from measurements of the magnetic coercivity $H{_C}(T)$.  An MH isotherm as measured at 2K for the Co${_2}$AlBO$_{5}$ system is shown in Figure \ref{figure7} .
 The fact that saturation is not reached till the highest applied magnetic fields indicates that at least some part of the sample persists in an antiferromagnetic phase right up to 7 Tesla. However, a striking observation is that the magnetization of Co${_2}$AlBO$_{5}$ is seen to exhibit sharp steps, exhibiting a staircase like behavior. Such sharp magnetization steps are now known to occur in a few strongly correlated systems, and are thought to arise from a number of distinct physical processes. For instance, in the spin chain compound Ca${_3}$Co${_2}$O${_6}$, well separated magnetization steps (and plateaus) are thought to arise from the quantum tunneling of magnetization \cite{maignan}. First postulated in the context of molecular magnets like Mn${_{12}}$-acetate \cite{mn12} and the Fe${_8}$ molecular nanomagnets \cite{Fe8}, this relies on the magnetic field facilitating resonant tunneling in systems characterized by a large spin values and an Ising-like anisotropy. Sharp magnetization steps have also been observed in site diluted antiferromagnets of the form Fe${_x}$Mg${_{1-x}}$Cl${_2}$ \cite{kleemann} and disordered systems like CeNi${_{1-x}}$Cu${_x}$ \cite{rfim1}, where they have been attributed to a field induced avalanche of flipping domains. Explained within the framework of the random field Ising model, these magnetization steps are typically characterized by the fact that their position appears to change in different measurements runs - no evidence of which is seen in our data.  

\begin{figure}
		\vspace*{-0.9cm}
		\hspace*{-1.3cm}
	\includegraphics[scale=0.36]{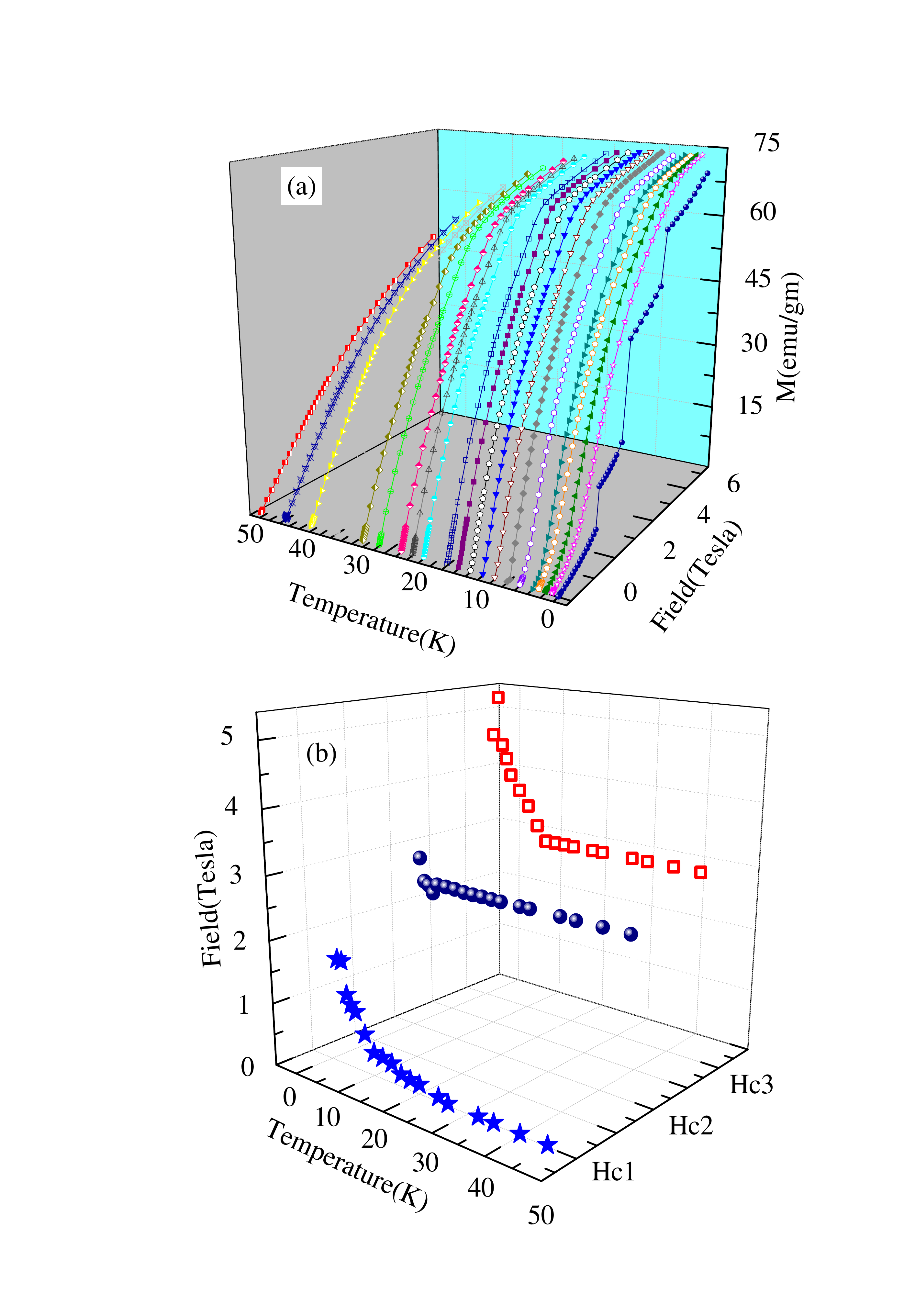}
	\caption{(a) depicts isotherms of the first magnetization curve of Co${_2}$AlBO$_{5}$ measured from 2 K to 40 K. The presence of 3 metamagnetic transitions can be deduced from $dM/dH$, and the critical fields associated with these transitions ($H{_{C1}}$, $H{_{C2}}$ and $H{_{C3}}$) are plotted in (b) as a function of temperature ($T$) and applied magnetic field ($H$). }
	\label{figure8}
\end{figure}

More recently, sharp magnetization steps have been reported in a number of transition metal oxides, where a martensitic scenario has been invoked. Gaining prominence in the context of phase separated manganites \cite{mahendiran, hardy, woodward,sunil1}, this scenario has also been used to explain magnetization steps in other material classes, including a recent report in an itinerant electron system LaFe${_{12}}$B${_6}$ \cite{new}. The mechanism here pertains to the catastrophic evolution of the antiferromagnetic (AFM) -ferromagnetic (FM) phase boundary when the applied magnetic field is used to change the predominantly AFM ground state to a FM one. The steps in magnetization are thus observed when the magnetization energy is minimized at the cost of the elastic energy at the AFM-FM interfaces. This was reminiscent of the growth of the martensite phase across an austenite-martensite phase transition, where the elastic strain at the interface between these two phases is relieved in sharp discontinuous steps. The glassy low temperature ground state of Co${_2}$AlBO$_{5}$ where both ferromagnetic and antiferromagnetic interactions are thought to co-exist could well belong to this class. We have also evaluated the evolution of the critical fields at which these magnetization steps occur, as a function of the temperature, as shown in Figure \ref{figure8}. Here, all the isotherms depict the first magnetization curve alone, and the system was heated to room temperatures between two consecutive isotherms. Traces of the three magnetization steps, which we observe in the MH isotherm at 2 K is also discernible in the form of a change of slope in the MH isotherms measured at higher temperatures. Using curves of $dM/dH$, we observe that these critical fields can be identified all throughout the magnetic phase, and is depicted in the lower panel of Figure \ref{figure8} .

It is evident that the critical fields associated with the metamagnetic transitions remain invariant of the measured temperatures right down to about 10 K, below which a monotonous increase of the critical field is observed. The fact that the magnetization has not saturated implies that additional metamagnetic transition(s) would possibly be observed at higher magnetic fields, before the system completely transforms to a ferromagnetic state. 

\section{Conclusions} 
In summary, we have investigated the structural, thermodynamic, magnetic and dielectric properties of the oxyborate Co${_2}$AlBO$_{5}$. This system is seen to crystallize in an orthorhombic $Pbam$ symmetry, and exhibits no evidence of a structural transition down to the lowest measured temperatures. This is in sharp contrast to the Fe-based ludwigite  Fe${_3}$BO${_5}$, which is characterized by a doubling of the unit cell along the crystallographic $c$ axis at $T \approx 280 K$, and highlights the difference between the Co and Fe based systems. On cooling,  Co${_2}$AlBO$_{5}$ exhibits an antiferro/ ferri-magnetic transition at 42K, followed by a superspin glass transition at $T{_G} = $10.6 K. Interestingly, a number of metamagnetic transitions are observed, which at the lowest temperatures manifest itself in the form of sharp magnetization steps, reminiscent of that observed in the mixed valent manganites. 

\section{Acknowledgements}
The authors thank D. Buddhikot and A. M. Patade for their help in heat capacity and dielectric measurements respectively. J.T acknowledges DST India for a SERB-NPDF. S.N. acknowledges DST India for support through grant no. SB/S2/CMP-048/2013. Authors thank the Department of Science and Technology, India for the financial support and Saha Institute of Nuclear Physics, India for facilitating the experiments at the Indian Beamline, Photon Factory, KEK, Japan.

 \bibliography{Bibliography}

\end{document}